\documentclass[10pt,conference]{IEEEtran}
\IEEEoverridecommandlockouts

\usepackage{multirow,makecell}
\usepackage[most]{tcolorbox}
\usepackage{color,xcolor}
\usepackage{listings,amsfonts}
\usepackage{caption}
\usepackage{subcaption}
\usepackage{threeparttable}
\usepackage{bbding}
\usepackage{graphicx}
\usepackage{fancyhdr}
\usepackage{booktabs}

\usepackage{pifont}
\usepackage{balance}

\usepackage{enumitem}
\usepackage[htt]{hyphenat}
\usepackage{array}

\newcounter{insightcounter}
\newtcolorbox{insightbox}[1][]{%
  enhanced,
  breakable,
  colback=gray!10,
  colframe=black,
  boxrule=0.5pt,
  arc=2pt,
  left=6pt,
  right=6pt,
  top=4pt,
  bottom=4pt,
  before upper={\stepcounter{insightcounter}\textbf{Insight \theinsightcounter: }\itshape}, 
  #1
}

\definecolor{customcite}{HTML}{e67a7a}
\definecolor{customlink}{HTML}{b83b5e}
\definecolor{customurl}{HTML}{11999e}
\AtEndPreamble{
\usepackage{hyperref}

    \hypersetup{
      colorlinks = true,
      linkcolor = customlink,
      anchorcolor = purple,
      citecolor = customcite,
      filecolor = purple,
      urlcolor = customurl
    }
}

\newcommand{\ea}{\textit{et~al.}}

\begin{document}

\title{Toward Understanding Bugs in Vector Database Management Systems}

\author{
\IEEEauthorblockN{Yinglin Xie\IEEEauthorrefmark{1}, Xinyi Hou\IEEEauthorrefmark{1}, Yanjie Zhao, Shenao Wang, Kai Chen\IEEEauthorrefmark{2} and Haoyu Wang\IEEEauthorrefmark{2}}
\IEEEauthorblockA{
Huazhong University of Science and Technology, Wuhan, China\\
\{xieyinglin, xinyihou, yanjie\_zhao, shenaowang, kchen, haoyuwang\}@hust.edu.cn}
\thanks{\IEEEauthorrefmark{1}Yinglin Xie and Xinyi Hou contributed equally to this work.}
\thanks{\IEEEauthorrefmark{2}The corresponding authors are Kai Chen (kchen@hust.edu.cn) and Haoyu Wang (haoyuwang@hust.edu.cn).}
}

\maketitle
\begin{abstract}
Vector database management systems (VDBMSs) play a crucial role in facilitating semantic similarity searches over high-dimensional embeddings from diverse data sources. While VDBMSs are widely used in applications such as recommendation, retrieval-augmented generation (RAG), and multimodal search, their reliability remains underexplored. Traditional database reliability models cannot be directly applied to VDBMSs because of fundamental differences in data representation, query mechanisms, and system architecture. To address this gap, we present the first large-scale empirical study of software defects in VDBMSs. We manually analyzed 1,671 bug-fix pull requests from 15 widely used open-source VDBMSs and developed a comprehensive taxonomy of bugs based on symptoms, root causes, and developer fix strategies. Our study identifies five categories of bug symptoms, with more than half manifesting as functional failures. We further reveal 31 recurring fault patterns and highlight failure modes unique to vector search systems. In addition, we summarize 12 common fix strategies, whose distribution underscores the critical importance of correct program logic. These findings provide actionable insights into VDBMS reliability challenges and offer guidance for building more robust future systems.
\end{abstract}

\section{Introduction}
\label{introduction}

Vector database management systems (VDBMSs) have emerged as a cornerstone of modern AI infrastructure, enabling the efficient management of high-dimensional embeddings and facilitating the semantic search across diverse data modalities. These systems underpin a wide range of transformative applications, including intelligent recommendation engines and cross-modal retrieval tasks such as using text queries to locate relevant video content~\cite{wu2025bhaktilightweightvectordatabase, yu2025mattersbridgingtopicssocial, Dong_2022}.

In the context of large language model (LLM) ecosystems, VDBMS plays a pivotal role in retrieval-augmented generation (RAG). Toolkits such as LangChain~\cite{Langchain} and LlamaIndex~\cite{LlamaIndex} integrate VDBMSs to ground LLM outputs with dynamically updated knowledge, thereby mitigating hallucination risks and improving factual consistency. Industry-leading systems like Vespa~\cite{Vespa}, Milvus~\cite{2021milvus}, and Faiss~\cite{douze2024faiss} have been carefully engineered to support heterogeneous hardware platforms, ranging from GPU clusters to neuromorphic processors. As demands for context-aware, low-latency LLM applications continue to grow, vector search capabilities are becoming a core priority within enterprise LLM stacks~\cite{8-Predictions,wang2025sok,wang2024llmsc}. 

As VDBMSs are increasingly used across a wide range of LLM applications, reliability is critically important. Software defects in these systems can result in serious consequences, such as corrupted vector indexes, inaccurate similarity computations, or even cascading failures in downstream LLM applications. For instance, inconsistencies in approximate nearest neighbor (ANN) algorithms can compromise recommendation accuracy in e-commerce platforms~\cite{wang2021-Survey}, while memory leaks in high-throughput environments may trigger system-wide instability in real-time scenarios~\cite{wang2025}. These issues not only diminish user trust but also impose substantial operational overhead in terms of debugging, mitigation, and recovery.

Traditional database management systems (DBMSs) have been extensively studied with well-established models for transaction processing and SQL optimization. However, these approaches are not directly applicable to VDBMSs due to fundamental architectural differences. VDBMSs differ from conventional systems in several key aspects. First, they operate on unstructured data by transforming it into high-dimensional numerical vectors, rather than storing structured records with fixed schemas. Second, their query mechanisms rely on probabilistic similarity search algorithms, such as ANN, rather than deterministic operations defined by SQL. Third, the semantic relationships among vectors give rise to dynamic and evolving data topologies that are incompatible with static schema constraints. These features challenge the assumptions of traditional reliability assessment techniques, making them insufficient for evaluating VDBMS behavior.

Despite their growing importance, there remains a significant gap in our understanding of the reliability risks and software defects specific to VDBMSs. To address this gap, we present the first large-scale empirical study of software defects in VDBMSs. We analyzed 1,671 bug-fix pull requests (PRs) from 15 widely used open-source VDBMSs hosted on GitHub. Each PR was carefully examined and categorized according to observable symptoms and root causes. Based on this analysis, we identified 5 symptom categories and 31 fault patterns, as well as 12 classes of fix strategies adopted by developers. This empirical approach allowed us to capture real-world failure scenarios, uncover recurring bug patterns and failure modes, and summarize typical repair strategies, ultimately offering actionable insights into the unique reliability challenges faced by VDBMSs. Our artifacts are publicly available at \url{https://figshare.com/s/00034c934612a54b8620}.

In summary, we make the following contributions:

\begin{itemize}
    \item \textbf{Empirical Analysis}: We conduct the first large-scale empirical study of software defects in VDBMSs, analyzing 1,671 bug-fix PRs from 15 open-source VDBMSs.
    
    \item \textbf{Taxonomy and Patterns}: We develop a comprehensive taxonomy of bugs based on symptoms, root causes, and fix strategies, and identify 5 symptom categories and 31 recurring fault patterns unique to VDBMSs.

    \item \textbf{Actionable Insights}: We uncover common failure modes and repair strategies, providing 10 actionable insights that inform both the development and testing of more reliable and robust VDBMSs.
\end{itemize}
\section{Background and Related Work}
\label{related work}

\subsection{Vector Database Management Systems (VDBMSs)}

\begin{figure}[htbp]
    \centering
    \includegraphics[width=0.95\linewidth]{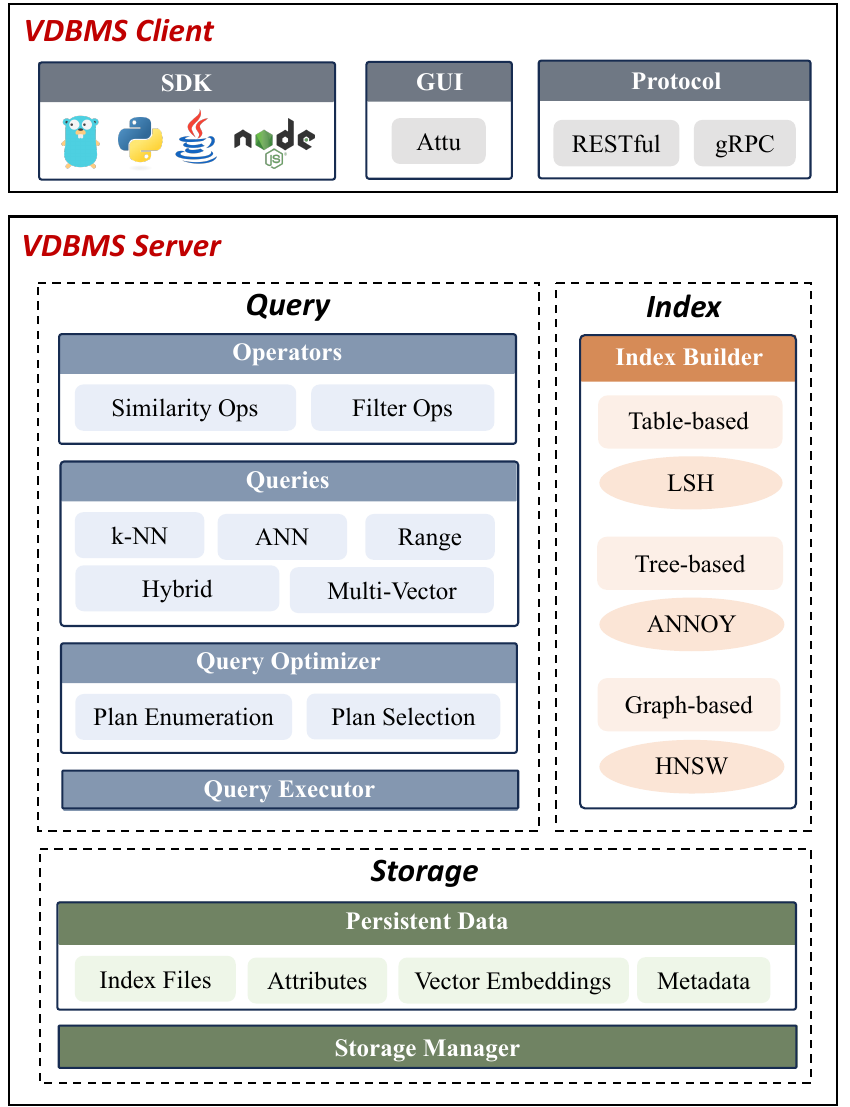} 
    \caption{Architecture of VDBMSs.}
    \label{fig:VDBMS}
\end{figure}

VDBMSs manage and retrieve high-dimensional vectors derived from unstructured data such as text, images, and audio. As shown in \autoref{fig:VDBMS}, they commonly adopt a client–server architecture: the server handles vector storage, indexing, and similarity search, while the client exposes interfaces for querying, data insertion, and metadata operations.

\begin{figure*}[t!]
    \centering
    \includegraphics[width=1\linewidth]{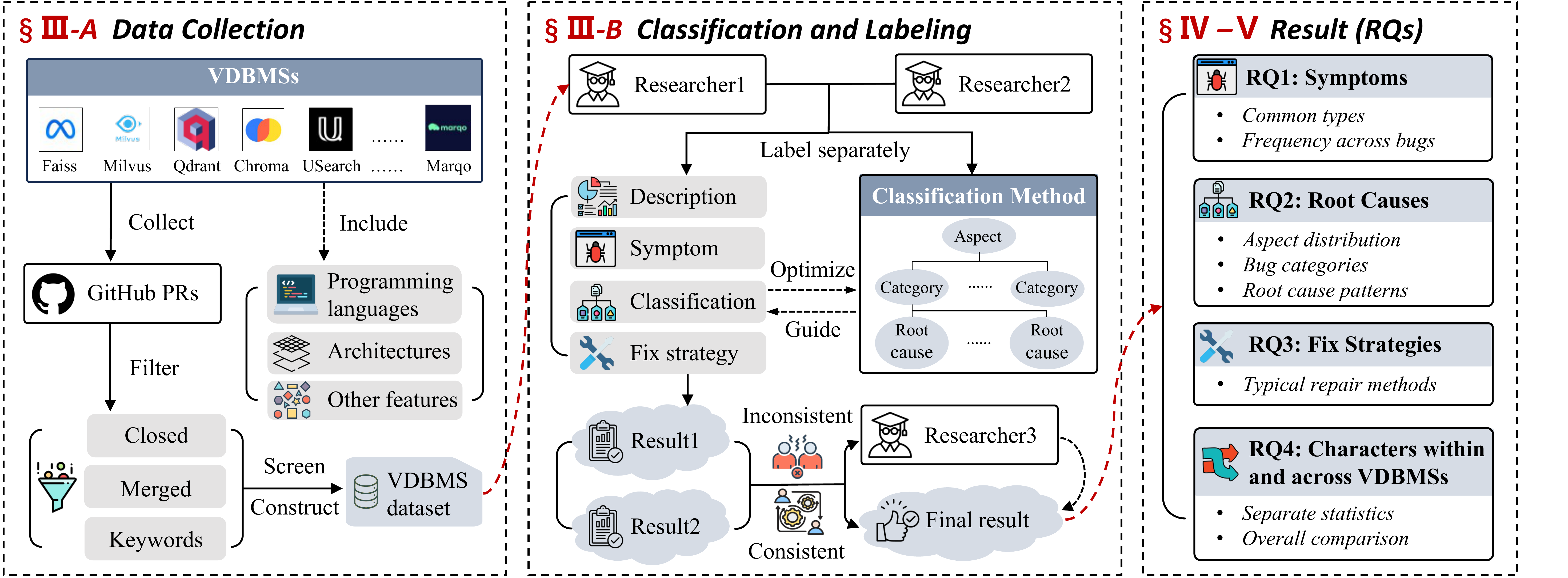} 
    \caption{Overview of the methodology for investigating bug characteristics in VDBMSs.}
    \label{fig:methodology}
\end{figure*}

\subsubsection{VDBMS Server}
User requests, transmitted via the client, flow through three layers on the server: query, index, and storage. The query layer parses and plans the request, the index layer performs efficient vector retrieval, and the storage layer provides access to the underlying data, which is then returned to the query layer for final processing and response.

\noindent\textbf{Query.} 
This layer includes four main components: operators, queries, the query optimizer, and the query executor. Operators perform tasks like similarity computation and result filtering, supporting query types such as k-nearest neighbor (k-NN)~\cite{Nearest-neighbor}, approximate search~\cite{ANN1993}, and range queries. For advanced use cases, it supports hybrid queries~\cite{wu2022hqann}, which combine vector similarity with attribute filters, and multi-vector queries, where multiple vectors and aggregated scores represent a single entity. The query optimizer selects efficient execution strategies, which are then carried out by the query executor through interaction with other VDBMS layers.

\noindent\textbf{Index.}  
The index layer is designed to accelerate similarity search over large collections of high-dimensional vectors by building specialized data structures that efficiently narrow down candidate results during query execution. Various indexing strategies can be employed, including table-based methods such as locality-sensitive hashing (LSH)~\cite{Muja2009}, tree-based approaches like random projection trees~\cite{NIPS2010_3def184a}, and graph-based structures such as HNSW~\cite{malkov2018efficient}.

\noindent\textbf{Storage.} 
This layer is responsible for managing all persistent data in the system, including vector embeddings, structured attributes, index files, and metadata. Its primary function is to ensure reliable storage and retrieval of this information while supporting efficient updates and access for query processing.

\subsubsection{VDBMS Client}
The client acts as the interface between users or applications and the VDBMS server. Most systems provide multi-language client SDKs, commonly written in Python, Java, Go, and JavaScript, which facilitate integration across diverse platforms. Some VDBMSs also offer graphical user interfaces to enhance client-side interaction; for example, Milvus provides a dedicated visual client called Attu~\cite{Attu}. Clients typically support two main communication protocols: REST APIs for lightweight metadata and control operations, and gRPC~\cite{grpc2025} for high-performance data transfer, particularly when handling large batches of vectors.

VDBMSs are essential for LLM applications using high-dimensional embeddings, providing core infrastructure for storing, searching, and managing vector data. They enable similarity-based retrieval for tasks like recommendation, anomaly detection, and semantic search, and also support RAG by supplying external knowledge to improve LLM responses.

\subsection{Reliability of VDBMSs}

The reliability of traditional DBMSs has long been a central topic of research. Liu~\ea~\cite{liu2024empiricalstudycharacteristicsdatabase} conducted a comprehensive analysis of 423 database access bugs across seven large-scale Java applications, while Cui~\ea~\cite{cui2024} examined 140 transaction-related bugs in six widely used database systems. Rigger~\ea~\cite{Rigger_2020} further introduced the Non-Optimizing Reference Engine Construction (NoREC) method to detect optimization bugs in query engines. These studies collectively reveal that DBMSs are susceptible to reliability issues stemming from diverse sources, including server-client interfacing, transaction handling, and query optimization.

In contrast, VDBMSs introduce a new set of reliability challenges due to their fundamental reliance on high-dimensional and semantically rich embeddings. Unlike traditional databases that operate over structured and ordinal data, VDBMSs must process unstructured data where query semantics are often ill-defined~\cite{tagliabue2023vectorspacefinalfrontier}. Moreover, the computational cost of vector comparisons is significantly higher, and building efficient yet consistent index structures for such data remains difficult~\cite{Wang2021,Wei2020,guo2022manucloudnativevector}. The complexity further increases with hybrid queries that integrate vector similarity with attribute-based filtering, complicating both query planning and execution. Although recent work has introduced various techniques to improve system efficiency, such as quantization-based compression~\cite{Gray,jegou2011,ite_matsui_2018} and learned partitioning methods~\cite{wang2017surveylearninghash,Muja2009,silpa2008}, these approaches largely focus on performance optimization. They offer limited solutions to deeper reliability concerns, such as incorrect results, inconsistent index states, or unpredictable behavior under dynamic workloads.
Given the growing use of VDBMSs in critical LLM applications, addressing their reliability challenges is both urgent and essential. This motivates our investigation into how these systems behave in practice and what factors underlie their failures.
\section{Methodology}
\label{methodology}

As shown in~\autoref{fig:methodology}, we designed a multi-step methodology involving data collection, classification, and labeling to systematically investigate bug characteristics in VDBMSs.

\subsection{Data Collection}

\subsubsection{VDBMS Selection}
We selected 15 VDBMSs as our research subjects, as listed in \autoref{tab:VecDBs}, including Faiss~\cite{douze2024faiss}, Milvus~\cite{2021milvus}, and Weaviate~\cite{Weaviate}. The selection followed several criteria: we prioritized widely adopted and actively maintained systems, using GitHub stars as a proxy for popularity, and included only those with over 14,000 stars as of November 29, 2024 to ensure broad usage and representativeness.
Horizontally, the selected VDBMSs span a range of implementation languages: Faiss~\cite{douze2024faiss}, Hnswlib~\cite{malkov2018efficient}, and Annoy~\cite{Annoy} are in C++; Qdrant~\cite{Qdrant} and pgvecto.rs~\cite{pgvecto.rs} in Rust; Chroma~\cite{Chroma}, txtai~\cite{txtai}, Deep Lake~\cite{deeplake}, and Voyager~\cite{Voyager} in Python; and Vespa~\cite{Vespa} in Java and C++. This diversity allows us to examine how language-level features affect bug characteristics. The systems also vary in architecture and indexing strategies, enabling broader analysis of how design choices impact reliability.
Vertically, the selected VDBMSs span a full decade, from 2013 to 2023, capturing the evolution of the field over time.  
The selection offers a diverse and representative foundation for our empirical analysis.

\begin{table}[h!]
\centering
\caption{Overview of Selected VDBMSs and Filtered PRs.}
\resizebox{\linewidth}{!}{ 
\begin{threeparttable}
\begin{tabular}{lcccrr}
\toprule[1.2pt]
\textbf{VDBMS} & \textbf{\#Forks} & \textbf{\#Stars} & \textbf{Language} & \textbf{\#Closed} & \textbf{\#Filtered} \\
\midrule[1.2pt]
Faiss~\cite{douze2024faiss}        & 3.8k  & 34.3k & C++             & 1,260   & 13    \\
Milvus~\cite{2021milvus}        & 3.1k  & 34.0k & Go, C++         & 23,392  & 412   \\
Qdrant~\cite{Qdrant}        & 1.6k  & 23.0k & Rust            & 3,939   & 35    \\
Chroma~\cite{Chroma}         & 1.6k  & 19.2k & Python          & 1,892   & 126   \\
Annoy~\cite{Annoy}         & 1.2k  & 13.7k & C++             & 263     & 2     \\
Weaviate~\cite{Weaviate}      & 0.9k  & 13.0k & Go              & 4,013   & 109   \\
txtai~\cite{txtai}         & 0.7k  & 10.7k & Python          & 38      & 8     \\
Deep Lake~\cite{deeplake}     & 0.6k  & 8.5k  & Python          & 2,478   & 167   \\
Vespa~\cite{Vespa}        & 0.6k  & 6.1k  & Java, C++       & 31,969  & 412   \\
LanceDB~\cite{LanceDB}       & 0.4k  & 6.1k  & Rust, Python    & 1,074   & 89    \\
Marqo~\cite{Marqo}         & 0.2k  & 4.8k  & Python          & 771     & 119   \\
Hnswlib~\cite{malkov2018efficient}       & 0.7k  & 4.6k  & C++             & 197     & 8     \\
Usearch~\cite{Vardanian_USearch_2023}       & 0.2k  & 2.6k  & C++             & 364     & 39    \\
pgvecto.rs~\cite{pgvecto.rs}    & 0.1k  & 2.0k  & Rust            & 341     & 115   \\
Voyager~\cite{Voyager}       & 0.1k  & 1.4k  & Python          & 63      & 17    \\
\midrule
\textbf{Total} & 15.8k & 184.0k & —              & 72,054  & 1,671 \\
\bottomrule[1.2pt]
\end{tabular}
\end{threeparttable}}
\label{tab:VecDBs}
\end{table}

\subsubsection{PR Collection}
We then sourced PRs from GitHub and applied a set of filtering criteria. We included only PRs that were \textbf{closed} and merged into the \textbf{main development branch}, which could be either \texttt{main} or \texttt{master}, depending on the project. To ensure relevance, we required that the PR title or labels contain \textbf{at least one} of the following keywords: \textit{bug}, \textit{error}, \textit{fail}, \textit{failure}, \textit{fault}, \textit{flaw}, \textit{mistake}, \textit{issue}, \textit{problem}, \textit{question}, \textit{matter}, \textit{trouble}, \textit{crash}, \textit{exception}, \textit{fix}, \textit{repair}, \textit{defect}, \textit{debug}, or \textit{correct}.
Milvus had an exceptionally large number of closed PRs (23,392). Due to the high cost of manual inspection, we selected a subset in reverse chronological order, covering August to December 2024, yielding 412 PRs. Vespa had a similar case, with 31,969 closed PRs. We adopted the same strategy and selected the most recent 412 PRs from December 2024.
Following the above steps, we obtained a comprehensive and representative dataset for our analysis. Detailed statistics, including the number of filtered PRs for each VDBMS, are provided in \autoref{tab:VecDBs}.

\subsection{Classification and Labeling}

\subsubsection{Taxonomy Construction}
\label{Taxonomy-Construction}
We collected 1,671 merged PRs from 15 widely used VDBMS projects. To enable systematic analysis, we labeled each PR using six key attributes: \textit{description}, \textit{symptom}, \textit{affected component}, \textit{root cause}, \textit{fix strategy}, and \textit{bug type}. These dimensions capture both observable symptoms and underlying causes of VDBMS bugs.
Among them, the \textbf{symptom} and \textbf{fix strategy} dimensions were adapted from prior studies on defect characterization~\cite{Quan_2022,chen2024understandingdeeplearningframework}. Symptoms are grouped into five categories (\autoref{fig:symptom}), while fix strategies are summarized into 12 recurring patterns (\autoref{tab:fix-strategies}), reflecting common failure modes and repair actions.
For the \textbf{root cause} dimension, we developed a hierarchical taxonomy grounded in the architecture of VDBMSs. Building on the system decomposition in~\cite{pan2024survey}, which separates a VDBMS into query processor and storage manager, we define five high-level categories: \textbf{Query}, \textbf{Storage}, \textbf{Index}, \textbf{Parsing \& Interaction}, and \textbf{Configuration}. Each category is further refined into subcategories and specific root causes.

\subsubsection{Annotation and Refinement}
We adopted a multi-stage collaborative procedure to annotate all 1,671 PRs using the taxonomy described above.
The process began with a pilot study, in which one researcher labeled a random sample of 150 PRs. During this phase, the root cause taxonomy was iteratively refined based on observed patterns, resulting in 20 initial root causes.
To assess annotation consistency, a second researcher independently annotated the same 150 PRs using the refined taxonomy. Inter-rater agreement, measured by Cohen’s Kappa~\cite{kappa}, exceeded 0.95, indicating near-perfect consistency. Disagreements were resolved by a third researcher serving as an arbitrator.
As annotation continued on the remaining PRs, new bug types emerged. These were either mapped to existing categories or temporarily assigned to a ``pending'' category. Through iterative discussion, the taxonomy was expanded to 41 root causes, and later consolidated into a final set of 31 to reduce redundancy and ambiguity.
Using the finalized taxonomy, symptom categories, and fix strategies, we re-annotated the dataset for consistency. Non-bug PRs were excluded, and those involving multiple distinct bugs were split into separate entries.  
After filtering, we obtained 1,463 confirmed bugs. The final taxonomy of root cause includes 5 top-level categories, 14 subcategories, and 31 leaf-level root causes. This high-quality, consistently labeled dataset serves as the basis for our subsequent analysis. 

\section{Taxonomy and Patterns}
\label{statistical results}

This section reports our empirical analysis of 1,463 real-world bug-fixing PRs in VDBMSs, examining symptoms, root causes, and fix strategies according to the following research questions (RQs).

\noindent\hangindent=2.5em\hangafter=1\textbf{RQ1 What are the typical symptoms exhibited by bugs in VDBMSs, and how are they distributed?}
We identify common bug symptoms in VDBMSs, analyze their frequency and diversity, and examine how they reflect different severity levels across categories.

\noindent\hangindent=2.5em\hangafter=1\textbf{RQ2 What are the most bug-prone components in VDBMSs, and what are the key root causes?}
We analyze the distribution of bugs across VDBMS components, categorize their types, and identify all root causes.

\noindent\hangindent=2.5em\hangafter=1\textbf{RQ3 What fix strategies are commonly adopted for different types of bugs in VDBMSs?}
We analyze how developers fix bugs by identifying common strategies and their association with different bug types.

\subsection{RQ1: Symptoms.}
\begin{figure}[h]
    \centering
    \includegraphics[width=1\linewidth]{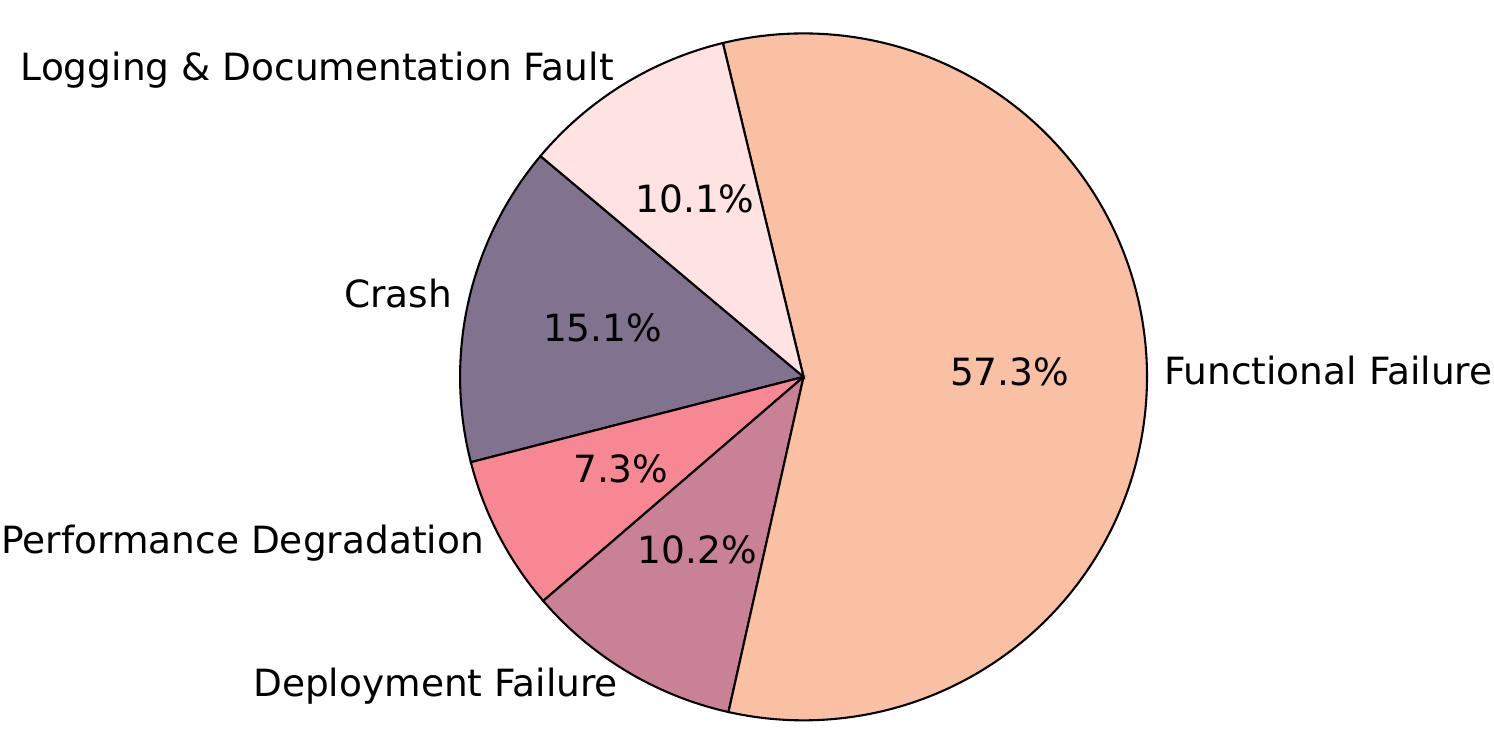}
    \caption{Distribution of symptoms.}
    \label{fig:symptom}
\end{figure}

Based on our observations, the symptoms were grouped into five major categories as shown in \autoref{fig:symptom}. These categories capture the diverse ways in which bugs can manifest, ranging from critical disruptions such as system crashes to less severe but still impactful issues like misleading log messages or incorrect documentation.

\subsubsection{Crash}

Crash is the most severe symptom observed in VDBMSs, as it directly causes service termination and loss of availability. Crash-related bugs account for 15.1\% of all symptoms in our dataset, with the highest proportions found in Qdrant (40.7\%) and txtai (40\%), indicating these systems are particularly prone to critical failures. In VDBMSs, crashes typically manifest as abrupt process terminations, segmentation faults, or unrecoverable runtime exceptions. These symptoms are frequently observed in core components such as storage backends, query pipelines, and system integration layers. Compared to traditional DBMSs, VDBMSs are more susceptible to crash bugs due to their reliance on high-performance native libraries (e.g., BLAS, Faiss) and complex vector algorithms, which increase the risk of memory errors and environment-dependent failures. The greater algorithmic complexity and varied deployment scenarios in VDBMSs also make edge-case crashes harder to test, reproduce, and diagnose.

\begin{insightbox}
To reduce crash risks, VDBMS developers should test edge cases, abnormal inputs (e.g., NaN or out-of-bound vectors), and concurrency. Components written in low-level languages like C or C++ require particular attention due to their vulnerability to memory errors.
\end{insightbox}

\subsubsection{Functional Failure}
This is the most common symptom in our dataset, accounting for 57.3\% of all cases. These bugs cause incorrect behaviors such as inaccurate query results, missing search matches, inconsistent indexing outputs, or invalid status responses. They often stem from logical inconsistencies or algorithmic flaws in query processing and indexing, which are highly complex and sensitive to semantic similarity definitions. For example, a functional bug in Faiss was triggered by misconfigured IVF search parameters, producing ``-1'' placeholder IDs in top-k results. Since vector queries rely on approximate similarity and complex indexes like quantized or graph-based ANN structures, result correctness is especially vulnerable to logic bugs in similarity projection or quantization. Fixing these issues typically requires major changes to logic or data flow.

\begin{insightbox}
Robust functional testing in VDBMSs should target both typical and edge-case query scenarios, with particular emphasis on validating the logic and parameter handling of approximate search and indexing algorithms.
\end{insightbox}

\subsubsection{Performance Degradation}
Performance bugs comprise 7.3\% of all cases, typically manifesting as excessive resource use, slow response times, or throughput bottlenecks. These often result from inefficient indexes, misconfigured parameters, or suboptimal query execution. For example, in Vespa (PR \#31,664), a linear-weight ranking profile with a high hits value caused nodes to process too many documents, significantly slowing queries. Unlike crashes or functional failures, performance degradation is usually gradual and workload-dependent.

\subsubsection{Deployment Failure}
Deployment failures in VDBMSs typically occur during build, installation, or startup phases. Common symptoms include missing dependencies, incompatible compiler versions, broken build scripts, and misconfigured environment variables. These issues account for 10.2\% of cases in our dataset and are often amplified by reliance on specialized native libraries (e.g., BLAS, Faiss, SIMD) and integration with containerization frameworks. Unlike runtime failures, deployment bugs can prevent system launch, posing a major barrier to adoption, especially for new users and in CI/CD environments. Robust and portable deployment processes are essential for improving VDBMS reliability.

\begin{insightbox}
VDBMS developers should ensure portability and environment isolation by using containerized builds, automated installation scripts, and pre-check tools to detect missing dependencies or incompatible settings.
\end{insightbox}

\subsubsection{Logging \& Documentation Fault}
This category accounts for 10.1\% of all recorded symptoms and includes missing, misleading, or outdated logs and documentation. Faulty logging may obscure root causes by omitting key runtime information, while poor documentation, such as incorrect usage examples or missing configuration guidance, hinders adoption and usage. For VDBMSs with complex queries and configurations, clear logs and up-to-date documentation are essential for troubleshooting and user onboarding.

\begin{figure*}[h!]
    \centering
    \begin{subfigure}{0.95\linewidth}
        \centering
        \includegraphics[width=\linewidth]{Figures/classification_query_2.pdf}
        \caption{Classification result for bugs in \textit{Query}.}
        \label{fig:classification-query}
    \end{subfigure}
    
    \begin{subfigure}{0.95\linewidth}
        \centering
        \includegraphics[width=\linewidth]{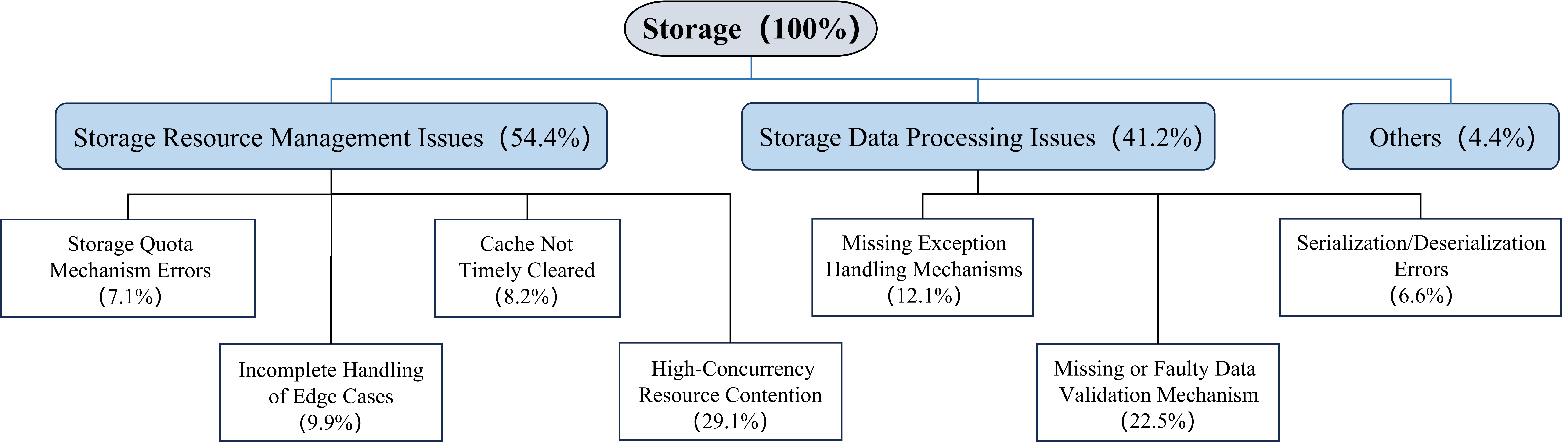}
        \caption{Classification result for bugs in \textit{Storage}.}
        \label{fig:classification-storage}
    \end{subfigure}
    
    \begin{subfigure}{0.95\linewidth}
        \centering
        \includegraphics[width=\linewidth]{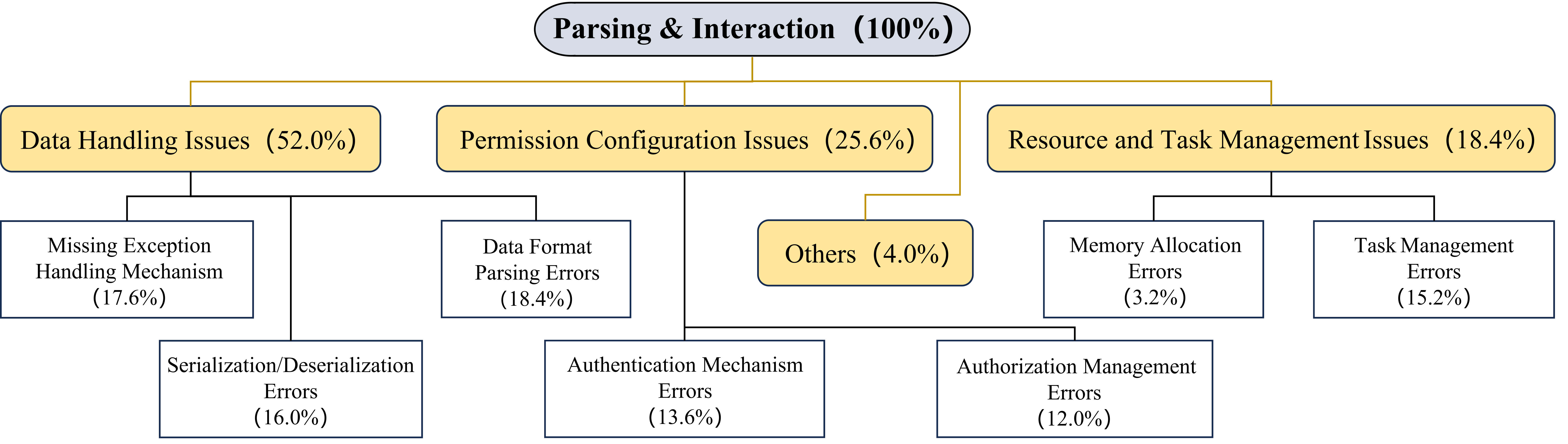}
        \caption{Classification result for bugs in \textit{Parsing \& Interaction}.}
        \label{fig:classification-parsing_interaction}
    \end{subfigure}

    \begin{subfigure}{0.65\linewidth}
        \centering
        \includegraphics[width=0.95\linewidth]{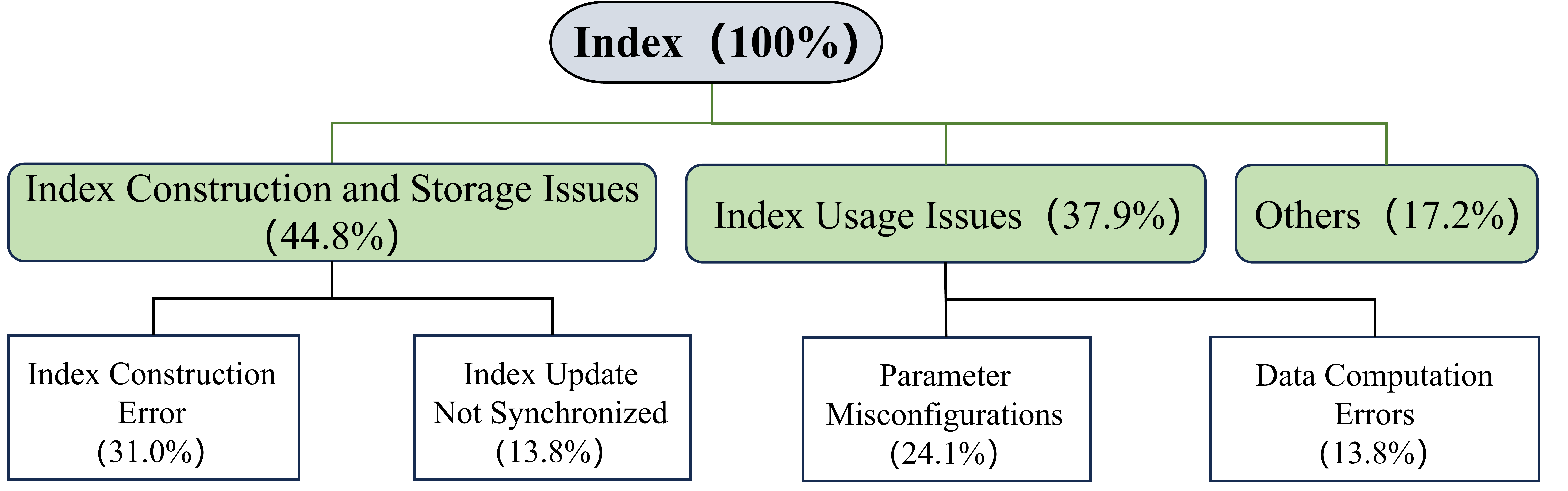}
        \caption{Classification result for Index.}
        \label{fig:classification-index}
    \end{subfigure}
    \hfill
    \begin{subfigure}{0.30\linewidth}
        \centering
        \includegraphics[width=0.9\linewidth]{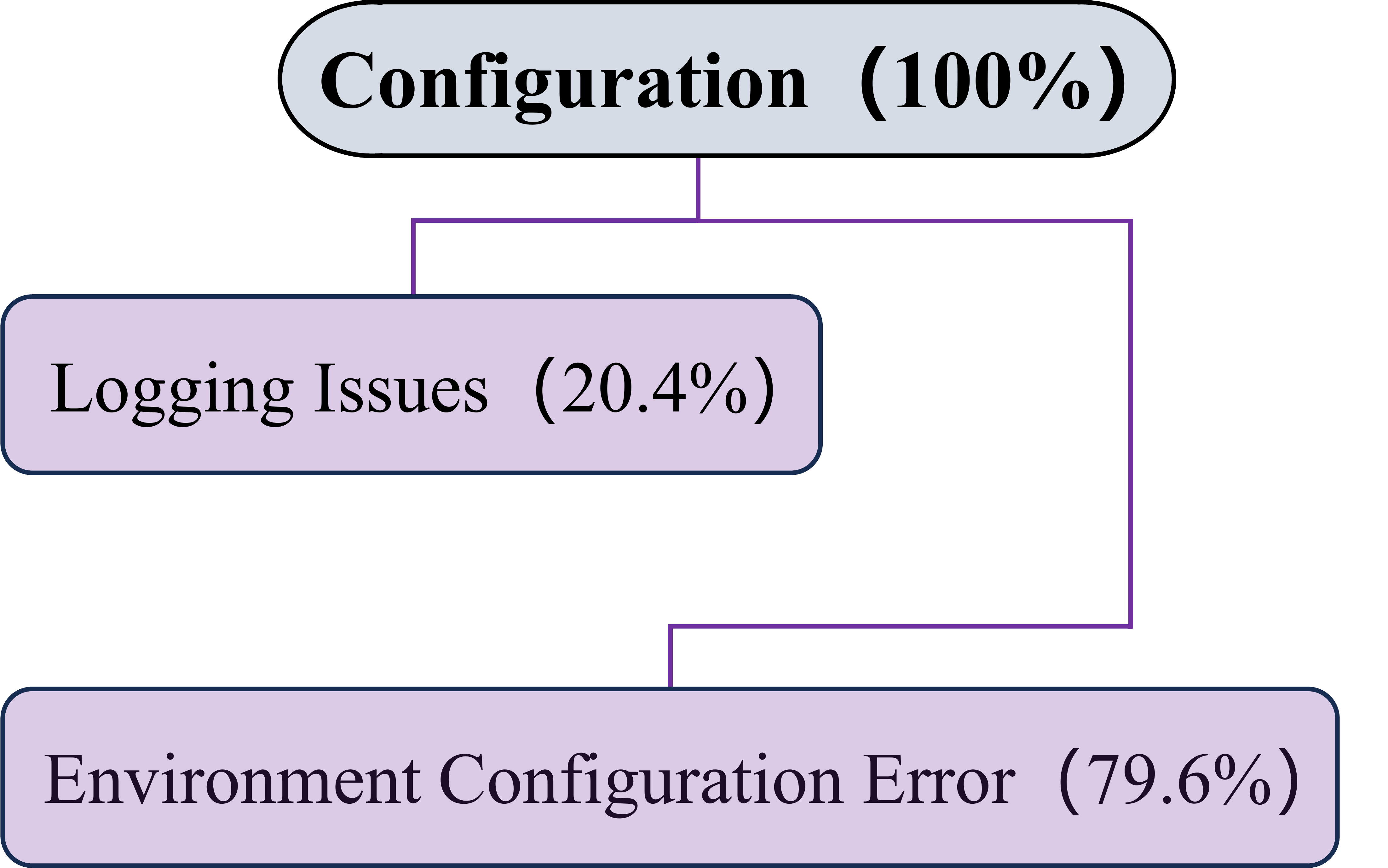}
        \caption{Classification result for Configuration.}
        \label{fig:classification-configuration}
    \end{subfigure}

    \caption{Overview and example of the bug classification method.}
    \label{fig:classification}
\end{figure*}
\subsection{RQ2: Root Causes.}
We construct a three-level taxonomy to classify bugs in VDBMSs, as shown in \autoref{fig:classification}. The percentages indicate the distribution of bugs at each level of the taxonomy. This taxonomy includes five categories: \textbf{Query} (711 bugs), \textbf{Storage} (182 bugs), \textbf{Index} (58 bugs), \textbf{Parsing \& Interaction} (125 bugs), and \textbf{Configuration} (387 bugs).

\subsubsection{Query} 
\label{subsub:query}
Query is the most bug-prone component, comprising 48.6\% of all cases. \textbf{\textit{Query algorithm logic issues}} (45.7\%) reflect challenges in ensuring correct query execution. \textbf{\textit{Query coordination issues}} (23.3\%) involve failures in managing concurrent execution and synchronization. \textbf{\textit{API interface issues}} (15.0\%) result from miscommunication between external APIs and internal handlers, while \textbf{\textit{memory management issues}} (8.7\%) stem from improper memory allocation or release during query processing.

\noindent\textbf{Query Algorithm Logic Issues.}
Incomplete or faulty logic in query algorithms, particularly when edge cases or unexpected inputs are overlooked, commonly leads to these bugs. \textbf{\textit{Missing exception handling mechanisms}}, such as failing to check for empty or null fields, which can lead to crashes. \textbf{\textit{Parameter misconfigurations}}, using illegal or ill-typed arguments, often result in subtle inaccuracies. \textbf{\textit{Data computation errors}}, like miscalculating similarity scores or buffer lengths, directly affect query reliability. \textbf{\textit{Filter abnormality}} refers to incomplete or incorrect filtering, causing improper inclusion or exclusion of data. These issues reflect the complexity of vector query logic and the challenge of anticipating all corner cases.

\begin{insightbox}
Ensuring the reliability of query algorithms in VDBMS requires thorough boundary testing and exception handling to address edge cases and unexpected parameter configurations.
\end{insightbox}

\noindent\textbf{Query Coordination Issues.}

Coordinating query stages or handling concurrent sessions can introduce subtle, hard-to-detect failures. One cause is \textbf{\textit{non-conformant data}}, where mismatched formats or inconsistent encoding disrupt execution. For example, in Weaviate (PR \#1,421), omitting the \texttt{\_additional} field led to incomplete downstream outputs. \textbf{\textit{State synchronization errors}} occur when components use outdated or inconsistent states, such as failing to terminate after a shutdown signal. \textbf{\textit{High-concurrency resource contention}} arises in parallel workloads, where poor thread or resource management results in race conditions, deadlocks, or degraded performance. Achieving reliable coordination requires consistent data interfaces, accurate state tracking, and careful concurrency control, which is particularly challenging in asynchronous and distributed VDBMS environments.

\noindent\textbf{API Interface Issues.}

System instability can arise from surface-level interfaces. \textbf{\textit{Parameter validation and handling errors}} occur when servers fail to verify input completeness, type, format, or range, or when dependencies and defaults are not properly enforced, leading to unexpected query behaviors. Another cause is \textbf{\textit{data transformation and response issues}}, often due to serialization or deserialization errors. For example, in Vespa (PR \#29,449), a misconfigured Jackson ObjectMapper resulted in malformed JSON during request parsing and response generation.
VDBMSs commonly process complex, nested queries with embeddings and hybrid parameters, where minor inconsistencies in input or response formatting can propagate system-wide. Robustness requires not just basic validation, but schema-aware design to ensure semantic consistency throughout the query lifecycle.

\noindent\textbf{Memory Management Issues.}  
High-dimensional vector computations introduce irregular, data-dependent memory access patterns that often lead to subtle memory errors.  
A common issue is \textbf{\textit{insufficient handling of boundary conditions}}, where the system fails to guard against edge cases such as buffer overflows or null pointer dereferences. For example, complex expressions evaluated over input vectors without proper checks can cause out-of-bounds access and crashes.  
Memory bugs also stem from \textbf{\textit{incorrect allocation strategies}}, including inaccurate memory estimation or poorly timed release. Premature deallocation can trigger use-after-free errors, while delayed release may cause memory leaks or bloat.  
In VDBMSs, memory usage is hard to predict due to factors like vector dimensionality, batch size, and filter complexity. To meet low-latency demands, systems often reuse buffers aggressively, increasing the risk of lifetime violations. These challenges are amplified by the large, high-dimensional nature of vector data, which is typically allocated in batches.

\begin{insightbox}
Memory allocation and usage in VDBMSs must account for the inherent complexity of vector data and various boundary conditions. The high dimensionality of vectors, coupled with data-dependent access patterns, demands precise memory estimation to avoid issues.
\end{insightbox}

\subsubsection{Storage}
As shown in \autoref{fig:classification-storage}, a total of 182 bugs are associated with the storage layer. These fall into two main categories: \textbf{\textit{Storage resource management issues}} (54.2\%), which concern the efficient utilization of storage resources and their impact on overall system performance; and \textbf{\textit{Storage data processing issues}} (41.2\%), which arise during data transformation, migration, or maintenance operations.

\noindent\textbf{Storage Data Processing Issues.}
VDBMSs are prone to data handling faults during storage operations, especially when serialization logic or validation mechanisms are poorly designed.  
\textbf{\textit{Serialization/Deserialization errors}} arise from mismatched formats, missing fields, or incompatible schema versions, leading to failures in persisting or restoring data. In some cases, valid data is rejected or causes crashes during deserialization.  
\textbf{\textit{Missing or faulty validation}} allows incorrect or corrupted records to enter the database, affecting downstream operations.  
\textbf{\textit{Missing exception handling}} can also hinder recovery from abnormal events. For example, in Milvus (PR \#36,149), the system failed to restore data after a crash, while in pgvecto.rs (PR \#242), a power loss left the database unrecoverable due to insufficient fault-tolerance logic.  
Unlike traditional DBMSs with well-defined schemas, VDBMSs often store high-dimensional data and index metadata in custom formats. This increases the risk of serialization mismatches, validation gaps, and deserialization errors, highlighting the need for specialized testing techniques.

\begin{insightbox}
Testing VDBMSs requires going beyond conventional strategies by incorporating various data schema fault injections to ensure robustness across heterogeneous and evolving data formats.
\end{insightbox}

\noindent\textbf{Storage Resource Management Issues.} 
Efficient storage resource utilization is critical in VDBMSs. 
In addition to the previously mentioned causes in \autoref{subsub:query}, such as \textbf{\textit{insufficient handling of boundary conditions}} and \textbf{\textit{high-concurrency resource contention}}, two other root causes are particularly relevant in this context. The first is \textbf{\textit{storage quota mechanism errors}}, where flawed strategies for allocating storage space, such as incorrect sharding logic, result in data being stored or retrieved incorrectly. The second is \textbf{cache not timely cleared}, where improper cache management, such as the failure to release memory after use, can lead to performance degradation or even system instability under sustained workloads.
Existing VDBMSs have adopted several practical strategies to optimize storage resource management, such as vector compression methods. However, our results show that bugs related to this category still account for 54.4\% of all storage-related issues. 

\begin{insightbox}
Stress testing VDBMS should focus on high-concurrency workloads, large-scale data ingestion, and edge cases like irregular sharding to ensure robust storage resource management under real-world conditions.
\end{insightbox}

\subsubsection{Parsing \& Interaction}
According to~\autoref{fig:classification-parsing_interaction}, 125 bugs are attributed to the parsing and interaction layer, stemming from client input parsing, result interpretation, and user-system interaction.
\textbf{\textit{Data handling issues}} (52\%), including failures in input parsing, format interpretation, and memory management during request and response processing.
\textbf{\textit{Permission configuration issues}} (25.6\%) involve authentication and access control, where misconfigurations result in failed or insecure interactions.  
\textbf{\textit{Resource and task management issues}} (18.4\%) concern incorrect memory allocation, poor task coordination, and errors in asynchronous execution.

\noindent\textbf{Data Handling Issues.}  
Data pipelines in VDBMSs are often fragile, breaking during request parsing or response reconstruction.  
A key root cause is \textbf{\textit{missing exception handling}}, where the system fails to detect or recover from malformed inputs, leading to crashes or undefined behavior.  
Another common issue is \textbf{\textit{data format parsing errors}}, caused by deviations from expected input structures, resulting in parsing failures or silent corruption.  
Response deserialization errors also occur when fields are missing, mismatched, or incompatible with the schema expected by the client.  
These failures highlight the need for defensive parsing strategies, such as schema validation, input fuzzing, and fallback mechanisms, to improve robustness against faulty inputs.

\noindent\textbf{Permission Configuration Issues.}
These issues mainly stem from \textbf{\textit{authentication mechanism errors}} or \textbf{\textit{authorization management errors}}. Authentication failures involve flawed user verification, allowing unauthorized access or blocking legitimate users. Authorization errors arise from misconfigured access rules, which can block valid actions or expose sensitive operations.
Such problems are critical in VDBMSs, which need fine-grained permissions for tasks like collection access, embedding uploads, and index changes. Operating in machine-to-machine environments or integrating with larger platforms, these systems are prone to subtle permission bugs from inconsistent roles, token scopes, or default settings.

\begin{insightbox}
Permission bugs in VDBMSs can block legitimate users or expose sensitive operations to unauthorized clients. Regular audits of configuration files and role definitions help detect misconfigurations early and ensure access controls align with intended security policies.
\end{insightbox}

\noindent\textbf{Resource and Task Management Issues.}  
The stability of VDBMSs often relies on careful control of memory usage and asynchronous task execution.
\textbf{\textit{Memory allocation errors}} occur when memory is allocated incorrectly, either too little or too much, during parsing or interaction, leading to performance issues or crashes. \textbf{\textit{Task management errors}} result from background operations that are not properly scheduled or monitored, causing inconsistent behavior.
For example, in LanceDB (PR \#407), a misconfigured timer led to irregular task execution and incorrect results. Robust VDBMSs require strict memory bounds, controlled asynchronous task lifecycles, and integrated observability mechanisms to detect anomalies early. These capabilities help prevent minor issues from propagating into major system failures.

\subsubsection{Index}
As shown in \autoref{fig:classification-index}, the Index aspect accounts for 58 bugs, representing only 4.0\% of the total. These are categorized into two types: \textbf{\textit{index construction and storage issues}}, which occur during index building, persistence, or updates; and \textbf{\textit{index usage issues}}, which arise during query execution using the index.
Despite their small number, indexing bugs can significantly affect system behavior. Structural or usage flaws may propagate to other components, resulting in incorrect queries or failures in data operations. As such, they often underlie issues in the query or storage layers and are critical to VDBMS robustness.

\noindent\textbf{Index Construction and Storage Issues.} 
Errors in building and maintaining index structures often lead to stale results or incomplete recall during search. 
Failures in this category often stem from \textbf{\textit{index construction error}}, where improper initialization, incorrect parameter configurations, or flawed data handling procedures result in incomplete or invalid index structures. Equally influential is \textbf{\textit{index update not synchronized}}, a condition in which updates to the underlying data are not correctly propagated to the index, leading to stale entries and inconsistent search outcomes.

\begin{insightbox}
For VDBMS testing, it is essential to validate that index structures are properly initialized, configured, and populated. Tests should also cover data updates and deletions to ensure index entries stay synchronized.
\end{insightbox}

\noindent\textbf{Index Usage Issues.} 
Bugs in this category typically reflect misuses of the index during query execution. Some problems stem from \textbf{\textit{parameter misconfigurations}}, where inappropriate settings for parameters such as search depth or probe count result in degraded accuracy or unnecessary performance overhead. Others involve \textbf{\textit{data computation errors}}, which can cause the system to return incomplete or misleading results. These issues do not compromise the structural integrity of the index but significantly undermine its utility during runtime.

\subsubsection{Configuration}
A total of 387 bugs (26.5\% of all cases) stem from configuration issues, making it one of the most significant sources of failure. As shown in \autoref{fig:classification-configuration}, a large portion involves \textbf{\textit{environment configuration errors}}, such as mismatches between documentation and actual API behavior, library or runtime incompatibilities, and incorrect dependency setups across deployment platforms.
Another cause is \textbf{\textit{logging-related problems}}, where missing, unclear, or misleading log messages obscure root causes and hinder debugging.
These bugs are particularly challenging due to their diffuse nature: they often arise not from a single module but from complex interactions among system components, external tools, and user environments. This highlights the need for holistic testing, tighter integration between code and tooling, and better alignment between documentation and actual behavior.

\subsection{RQ3: Fix Strategies.}
We identify 12 representative fix strategies from an analysis of VDBMS bug-fixing commits. Each strategy reflects how developers resolve bugs by targeting configuration, code logic, data handling, or environment settings. \autoref{tab:fix-strategies} summarizes their distribution across four key areas: \textbf{Environment}, \textbf{Data}, \textbf{Program}, and \textbf{API}. We further group and analyze these strategies to understand their frequency and use cases.

\begin{table}[h]
\centering
\caption{Fix strategies adopted across four system layers.}
\resizebox{\linewidth}{!}{
\begin{threeparttable}
\begin{tabular}{clr}
\toprule[1.2pt]
\textbf{Object} & \textbf{Fix Strategy} & \textbf{Count} \\
\midrule[1.2pt]
\multirow{4}{*}{Environment}
& Upgrade version                          & 30 \\
& Modifying dependency configurations      & 34 \\
& Modify environment variable values       & 41 \\
& Update configuration files or instructions & 134 \\
\midrule
\multirow{2}{*}{Data}
& Modify or add data processing logic      & 40 \\
& Modify data default values or initialization & 46 \\
\midrule
\multirow{5}{*}{Program}
& Fix or complete program logic            & 706 \\
& Modify parameter values                  & 54 \\
& Adjust memory allocation strategy        & 71 \\
& Add validation and safeguard mechanisms  & 188 \\
& Remove redundant logic                   & 60 \\
\midrule
API
& Adjust API usage                         & 45 \\
\midrule
Others
&  Model modification, Error message update…        & 14 \\
\midrule
\textbf{Total} &  & \textbf{1,463} \\
\bottomrule[1.2pt]
\end{tabular}
\end{threeparttable}}
\label{tab:fix-strategies}
\end{table}

\begin{table*}[h!]
\centering
\caption{Distribution of bug symptoms and the aspects of taxonomy across 15 vector databases.}
\resizebox{0.95\linewidth}{!}{
\begin{threeparttable}
\begin{tabular}{lrrrrrrrrrrrr}
\toprule[1.2pt]
\textbf{VecDB} & \multicolumn{5}{c}{\textbf{Symptom (\%)}} & \multicolumn{5}{c}{\textbf{Taxonomy (Aspect) (\%)}} & \textbf{Total} \\
\cmidrule(lr){2-6} \cmidrule(lr){7-11}
& \textbf{Crash} & \textbf{Perf Degrad.} & \textbf{Deploy Fail.} & \textbf{Func Error} & \textbf{Log/Doc} & \textbf{Query} & \textbf{Storage} & \textbf{Index} & \textbf{Parse/Inter.} & \textbf{Config} & \\
\midrule
Faiss       & 11.1\% & / & 77.8\% & 11.1\% & / & / & 11.1\% & 11.1\% & / & 77.8\% & 9 \\
Milvus      & 17.1\% & 11.6\% & 4.4\% & 62.3\% & 4.7\% & 65.0\% & 18.2\% & 6.6\% & 1.7\% & 8.5\% & 363 \\
Qdrant      & 40.7\% & 11.1\% & 18.5\% & 22.2\% & 7.4\% & 18.5\% & 40.7\% & 3.7\% & / & 37.0\% & 27 \\
Chroma      & 17.2\% & 2.3\% & 23.0\% & 55.2\% & 2.3\% & 29.0\% & 15.1\% & 4.3\% & 7.5\% & 44.1\% & 93 \\
Annoy       & / & / & / & / & 100.0\% & / & / & 100.0\% & / & / & 1 \\
Weaviate    & 28.3\% & 6.5\% & 6.5\% & 53.3\% & 5.4\% & 42.4\% & 31.5\% & 4.3\% & 3.3\% & 18.5\% & 92 \\
txtai       & 40.0\% & 20.0\% & / & / & 40.0\% & 20.0\% & / & / & / & 80.0\% & 5 \\
Deep Lake   & 10.7\% & 4.8\% & 6.0\% & 66.7\% & 11.9\% & 45.3\% & 13.0\% & 0.6\% & 19.3\% & 21.7\% & 168 \\
Vespa       & 6.6\% & 8.9\% & 9.5\% & 62.2\% & 12.7\% & 52.6\% & 4.9\% & 1.1\% & 10.3\% & 31.0\% & 348 \\
LanceDB     & 17.1\% & 3.9\% & 15.8\% & 47.4\% & 15.8\% & 13.2\% & 14.5\% & 5.3\% & 15.8\% & 51.3\% & 76 \\
Marqo       & 7.5\% & 2.8\% & 8.4\% & 63.6\% & 17.8\% & 50.5\% & 6.5\% & 3.7\% & 2.8\% & 36.4\% & 107 \\
Hnswlib     & 33.3\% & / & 16.6\% & 33.3\% & 16.6\% & 33.3\% & / & 16.6\% & / & 50.0\% & 6 \\
Usearch     & 21.1\% & 2.6\% & 21.1\% & 44.7\% & 10.5\% & 50.0\% & / & 2.6\% & 10.5\% & 36.8\% & 38 \\
pgvecto.rs  & 13.4\% & 8.0\% & 13.4\% & 52.7\% & 12.5\% & 46.9\% & 4.4\% & 6.2\% & 2.7\% & 39.8\% & 113 \\
Voyager     & 11.8\% & / & 47.1\% & 41.2\% & / & 23.5\% & / & 5.9\% & 11.8\% & 58.8\% & 17 \\
\bottomrule[1.2pt]
\end{tabular}
\end{threeparttable}}
\label{tab:rq4-1}
\end{table*}

\subsubsection{Environment} 
Fix strategies targeting the environment layer account for a substantial portion of all observed cases (239 in total). These are mainly used to resolve misconfigurations, compatibility issues, or incorrect setup procedures. The most common strategy is \textit{\textbf{updating configuration files or instructional materials}} (134 cases), which includes correcting default templates, adding missing setup steps, or clarifying deployment guides. Another frequent approach is \textit{\textbf{modifying environment variable values}} (41 cases) to fix path errors, resource visibility issues, or to enable or disable specific runtime behaviors. \textit{\textbf{Upgrading versions}} (30 cases) typically addresses library incompatibilities, outdated dependencies, or deprecation warnings. Finally, \textit{\textbf{modifying dependency configurations}} (34 cases) involves adding missing packages, adjusting version constraints, or removing conflicting dependencies. These changes are usually made through files such as \texttt{requirements.txt}, \texttt{poetry.lock}, or Dockerfiles.

\subsubsection{Data} 
Fix strategies at the data layer mainly fall into two categories. The first involves \textit{\textbf{modifying or adding data processing logic}} (40 cases), such as correcting preprocessing errors or adding missing normalization steps in ingestion pipelines. The second addresses \textit{\textbf{modifying data default values or initialization}} (46 cases), which often includes adjusting placeholder values, setting appropriate vector dimensions, or ensuring proper initialization of data fields to prevent runtime exceptions or incorrect computations.

\subsubsection{Program}
Program-level fix strategies are the most prevalent, addressing 1,079 bugs by modifying core logic to fix semantic errors, complete missing functionality, and handle edge cases. \textit{\textbf{Fixing or completing program logic}} (706 cases), which includes correcting control flow, adding missing conditional branches, or fixing flawed computations. Developers also often \textit{\textbf{add validation and safeguard mechanisms}} (188 cases) to prevent runtime exceptions and enforce correctness, such as checks for null values, vector dimensions, or operation preconditions. Other strategies include \textit{\textbf{removing redundant logic}} (60 cases), often left over from earlier development or refactoring, and \textit{\textbf{modifying parameter values}} (54 cases) to adjust thresholds, batch sizes, or resource limits. In performance-critical paths, \textit{\textbf{adjusting memory allocation strategies}} (71 cases) helps avoid out-of-memory errors and improve efficiency, particularly during large-scale vector operations.

\begin{insightbox}
Many program-level bugs in VDBMS stem from logic and edge-case errors that are hard to detect with standard unit tests. 
\end{insightbox}

\subsubsection{API}
Bugs related to incorrect or outdated API usage are addressed through a single but important strategy: \textit{\textbf{adjusting API usage}} (45 cases). These fixes typically involve modifying invocation methods, correcting argument formats, or reordering dependent function calls. Such issues are often triggered by breaking changes in upstream libraries or misuse of internal interfaces within the VDBMS engine.
To mitigate such issues, it is essential to strengthen interface contracts and improve change management practices. Effective measures include enforcing version checks, monitoring deprecation warnings, and isolating internal APIs from external exposure, especially in fast-evolving codebases.

\section{Correlation and Cross-System Analysis}
\label{Correlation and Cross-System Analysis}

This section explores bug characteristics within and across 15 VDBMSs, guided by the following RQ:

\noindent\hangindent=2.5em\hangafter=1\textbf{RQ4 How do symptoms, root causes, and fix strategies differ across VDBMSs?}
Our goal is to understand how variations in bug manifestation and resolution across VDBMSs relate to system-specific factors such as architectural design or programming language.

\subsection{RQ4: Analysis across VDBMSs.}

\subsubsection{Bug Symptoms Reflect Architectural Design}
As shown in \autoref{tab:rq4-1}, functional failures are the most common bug symptom across the evaluated systems. This pattern is especially pronounced in native VDBMSs such as Milvus (62.3\%) and Chroma (55.2\%), where performance-critical query execution pipelines are built from scratch. These systems often lack mature query engines, making them more prone to execution errors.
In contrast, extended VDBMSs that build on top of existing DBMS infrastructures tend to exhibit a broader range of symptoms, including deployment and configuration failures. Their reliance on stable components reduces deep execution bugs, but increases the likelihood of integration-related issues.
Qdrant is a notable outlier, with 40.7\% of its bugs located in the storage layer. This can be attributed to its architectural decision to implement a custom storage backend, including its write-ahead log, snapshotting, and segment persistence mechanisms.
These patterns suggest that \textbf{bug symptoms tend to concentrate in areas where systems deviate most from standardized infrastructure}, highlighting the cost of custom architecture in terms of reliability risks.

\subsubsection{Query-Related Bugs Vary with Functional Scope}
A notable difference across VDBMSs lies in the query layer. Mostly-vector systems, such as Chroma, focus on fast approximate nearest neighbor search with limited query complexity. These systems usually support a single index type and fixed execution flow, avoiding query optimizers or planners. As a result, query-related bugs are relatively rare (29.0\%).
In contrast, mostly-mixed systems like Milvus (65.0\%) and Marqo (50.5\%) support more expressive queries, including filtering, hybrid search, and multiple index types. These features introduce additional complexity and larger interaction surfaces between components such as planners, optimizers, and indexing modules, increasing the likelihood of query-related failures.
\textbf{Richer query functionality expands retrieval power but also increases the risk of bugs}, especially when query logic interacts with multiple tightly coupled modules.

\subsubsection{Fix Strategies Reveal Differences in Modularity and Runtime Environment}
Distribution of fix strategies across VDBMSs\footnote{Detailed distributions of fix strategies across 15 VDBMSs are available in our supplementary artifacts at \url{https://figshare.com/s/00034c934612a54b8620}.} shows notable differences between systems. Milvus resolves 59.8\% of its bugs through direct logic modifications, reflecting its monolithic and execution-centric design. 
Chroma demonstrates a more varied pattern, with 45.2\% logic fixes, but also more dependency (14.0\%) and environment (9.7\%) changes, likely due to its Python-based implementation and ML toolchain integration. 
Systems like txtai, LanceDB, and Vespa rely more on configuration and environment-level fixes, indicating greater use of flexible deployments and scripting, which influence both failures and fix patterns.

\section{Discussion}
\label{discussion}

\subsection{Implications}
Our findings offer practical guidance for VDBMS developers, architects, and researchers to improve reliability and robustness in real-world systems.
\subsubsection{For VDBMS Developers.}
\textbf{\textit{Prioritize logic robustness in query modules.}} Query Algorithm Logic Issues are the most common root cause (45.7\%, see \autoref{fig:classification-query}), often due to missing exception handling or incorrect similarity computations. Developers should strengthen defensive logic, particularly for parameter validation and boundary conditions. 
\textit{\textbf{Address concurrency and memory management.}} Issues like state synchronization failures and memory mismanagement (see \autoref{fig:classification-storage}) highlight the need for better resource control under high-concurrency workloads. Using concurrency-safe data structures and finer-grained lifecycle management can help reduce these bugs. 
\textbf{\textit{Expand automated test coverage for edge cases.}} Since many logic bugs are triggered by rare inputs (e.g., empty vectors, null IDs), developers should systematically adopt automated test suites to cover these scenarios.

\subsubsection{For VDBMS Architects.} 
\textbf{\textit{Ensure configuration is robust and introspectable.}} Configuration bugs account for 26.5\% of cases (see \autoref{fig:classification-configuration}), such as missing environment variables and incompatible dependencies. We recommend startup validation, machine-checkable formats, and self-reporting diagnostics to catch misconfigurations.
\textit{\textbf{Emphasize unit testing.}} Incorrect code logic is the main root cause of hard-to-trace failures. Architects should ensure key modules like index updates and query coordination are thoroughly tested with both normal and edge-case inputs.

\subsubsection{For VDBMS Researchers.} 
\textbf{\textit{Explore memory-safe and concurrency-resilient execution frameworks.}} Many crashes are caused by memory management and query coordination issues, such as use-after-free, state desynchronization, and high-concurrency contention. Researchers should investigate runtime systems or execution models optimized for bursty, data-dependent query workloads. 
\textbf{\textit{Advance automated bug detection.}} Functional failures are the most common symptom (57.3\%) and are difficult to detect with traditional analysis. Future work should develop domain-specific verification or fuzzing tools for vector quantization and similar operations.

\subsection{Limitations}

\noindent\textbf{Granularity of taxonomy.}  
Given the diversity and complexity of VDBMS bugs, our taxonomy may not capture all fine-grained distinctions. To address this, we used an iterative annotation process, refining our classification criteria in parallel with the labeling. When updates were made, previously labeled cases were revisited and adjusted to ensure consistency.

\noindent\textbf{Manual labeling.}  
Our classification of bug symptoms, root causes, and fix strategies relied on manual labeling, which inevitably introduces some subjectivity. To mitigate this, two researchers independently labeled each sample, and disagreements were resolved by a third domain expert through discussion. Ultimately, the consensus was reached on all cases.

\section{Conclusion}
\label{conclusion}

VDBMSs are essential to modern AI applications, yet their reliability is not well understood. This study presents the first large-scale empirical analysis of VDBMS bugs, based on 1,671 bug-fix pull requests from 15 open-source systems. We developed a taxonomy of bugs covering symptoms, root causes, and fix strategies, identifying five major symptom categories, 31 recurring fault patterns, and 12 common fix strategies. Our findings highlight unique failure modes in vector search workloads and emphasize the central role of program logic in bug fixes. These insights offer practical guidance for improving the robustness of future VDBMSs.

\balance
\bibliographystyle{IEEEtranS}
\bibliography{main}

\begin{thebibliography}{10}
\providecommand{\url}[1]{#1}
\csname url@samestyle\endcsname
\providecommand{\newblock}{\relax}
\providecommand{\bibinfo}[2]{#2}
\providecommand{\BIBentrySTDinterwordspacing}{\spaceskip=0pt\relax}
\providecommand{\BIBentryALTinterwordstretchfactor}{4}
\providecommand{\BIBentryALTinterwordspacing}{\spaceskip=\fontdimen2\font plus
\BIBentryALTinterwordstretchfactor\fontdimen3\font minus \fontdimen4\font\relax}
\providecommand{\BIBforeignlanguage}[2]{{%
\expandafter\ifx\csname l@#1\endcsname\relax
\typeout{** WARNING: IEEEtranS.bst: No hyphenation pattern has been}%
\typeout{** loaded for the language `#1'. Using the pattern for}%
\typeout{** the default language instead.}%
\else
\language=\csname l@#1\endcsname
\fi
#2}}
\providecommand{\BIBdecl}{\relax}
\BIBdecl

\bibitem{Annoy}
Annoy, ``Annoy,'' \url{https://github.com/spotify/annoy}, 2013.

\bibitem{ANN1993}
S.~Arya and D.~M. Mount, ``Approximate nearest neighbor queries in fixed dimensions,'' in \emph{Proceedings of the Fourth Annual ACM-SIAM Symposium on Discrete Algorithms}, ser. SODA '93.\hskip 1em plus 0.5em minus 0.4em\relax USA: Society for Industrial and Applied Mathematics, 1993, p. 271–280.

\bibitem{Attu}
{Attu}, ``{Attu},'' \url{https://zilliz.com.cn/attu}, 2022.

\bibitem{chen2024understandingdeeplearningframework}
\BIBentryALTinterwordspacing
J.~Chen, Y.~Liang, Q.~Shen, J.~Jiang, and S.~Li, ``Toward understanding deep learning framework bugs,'' 2024. [Online]. Available: \url{https://arxiv.org/abs/2203.04026}
\BIBentrySTDinterwordspacing

\bibitem{Chroma}
Chroma, ``Chroma,'' \url{https://github.com/chroma-core/chroma}, 2022.

\bibitem{Nearest-neighbor}
T.~Cover and P.~Hart, ``Nearest neighbor pattern classification,'' \emph{IEEE Transactions on Information Theory}, vol.~13, no.~1, pp. 21--27, 1967.

\bibitem{cui2024}
\BIBentryALTinterwordspacing
Z.~Cui, W.~Dou, Y.~Gao, D.~Wang, J.~Song, Y.~Zheng, T.~Wang, R.~Yang, K.~Xu, Y.~Hu, J.~Wei, and T.~Huang, ``Understanding transaction bugs in database systems,'' in \emph{Proceedings of the IEEE/ACM 46th International Conference on Software Engineering}, ser. ICSE '24.\hskip 1em plus 0.5em minus 0.4em\relax New York, NY, USA: Association for Computing Machinery, 2024. [Online]. Available: \url{https://doi.org/10.1145/3597503.3639207}
\BIBentrySTDinterwordspacing

\bibitem{NIPS2010_3def184a}
\BIBentryALTinterwordspacing
A.~Dhesi and P.~Kar, ``Random projection trees revisited,'' in \emph{Advances in Neural Information Processing Systems}, J.~Lafferty, C.~Williams, J.~Shawe-Taylor, R.~Zemel, and A.~Culotta, Eds., vol.~23.\hskip 1em plus 0.5em minus 0.4em\relax Curran Associates, Inc., 2010. [Online]. Available: \url{https://proceedings.neurips.cc/paper_files/paper/2010/file/3def184ad8f4755ff269862ea77393dd-Paper.pdf}
\BIBentrySTDinterwordspacing

\bibitem{Dong_2022}
\BIBentryALTinterwordspacing
J.~Dong, X.~Chen, M.~Zhang, X.~Yang, S.~Chen, X.~Li, and X.~Wang, ``Partially relevant video retrieval,'' in \emph{Proceedings of the 30th ACM International Conference on Multimedia}, ser. MM ’22.\hskip 1em plus 0.5em minus 0.4em\relax ACM, Oct. 2022, p. 246–257. [Online]. Available: \url{http://dx.doi.org/10.1145/3503161.3547976}
\BIBentrySTDinterwordspacing

\bibitem{douze2024faiss}
M.~Douze, A.~Guzhva, C.~Deng, J.~Johnson, G.~Szilvasy, P.-E. Mazaré, M.~Lomeli, L.~Hosseini, and H.~Jégou, ``The faiss library,'' 2024.

\bibitem{Gray}
R.~Gray, ``Vector quantization,'' \emph{IEEE ASSP Magazine}, vol.~1, no.~2, pp. 4--29, 1984.

\bibitem{grpc2025}
{gRPC}, ``{gRPC},'' \url{https://github.com/grpc/grpc}, 2025.

\bibitem{guo2022manucloudnativevector}
\BIBentryALTinterwordspacing
R.~Guo, X.~Luan, L.~Xiang, X.~Yan, X.~Yi, J.~Luo, Q.~Cheng, W.~Xu, J.~Luo, F.~Liu, Z.~Cao, Y.~Qiao, T.~Wang, B.~Tang, and C.~Xie, ``Manu: A cloud native vector database management system,'' 2022. [Online]. Available: \url{https://arxiv.org/abs/2206.13843}
\BIBentrySTDinterwordspacing

\bibitem{deeplake}
\BIBentryALTinterwordspacing
S.~Hambardzumyan, A.~Tuli, L.~Ghukasyan, F.~Rahman, H.~Topchyan, D.~Isayan, M.~Harutyunyan, T.~Hakobyan, I.~Stranic, and D.~Buniatyan, ``Deep lake: a lakehouse for deep learning,'' 2023. [Online]. Available: \url{https://www.cidrdb.org/cidr2023/papers/p69-buniatyan.pdf}
\BIBentrySTDinterwordspacing

\bibitem{jegou2011}
H.~Jégou, M.~Douze, and C.~Schmid, ``Product quantization for nearest neighbor search,'' \emph{IEEE Transactions on Pattern Analysis and Machine Intelligence}, vol.~33, no.~1, pp. 117--128, 2011.

\bibitem{LanceDB}
LanceDB, ``Lancedb,'' \url{https://github.com/lancedb/lancedb}, 2023.

\bibitem{Langchain}
LangChain, ``Langchain,'' \url{https://www.langchain.com/}, 2022.

\bibitem{liu2024empiricalstudycharacteristicsdatabase}
\BIBentryALTinterwordspacing
W.~Liu, S.~Mondal, and T.-H. Chen, ``An empirical study on the characteristics of database access bugs in java applications,'' 2024. [Online]. Available: \url{https://arxiv.org/abs/2405.15008}
\BIBentrySTDinterwordspacing

\bibitem{LlamaIndex}
LlamaIndex, ``Llamaindex,'' \url{https://docs.llamaindex.ai/}, 2023.

\bibitem{8-Predictions}
J.~Luan, ``The next stop for vector databases: 8 predictions for 2023,'' \url{https://zilliz.com/blog/the-next-stop-for-vector-databases-8-predictions-for-2023}, 2022.

\bibitem{malkov2018efficient}
Y.~A. Malkov and D.~A. Yashunin, ``Efficient and robust approximate nearest neighbor search using hierarchical navigable small world graphs,'' \emph{IEEE transactions on pattern analysis and machine intelligence}, vol.~42, no.~4, pp. 824--836, 2018.

\bibitem{Marqo}
Marqo, ``Marqo,'' \url{https://github.com/marqo-ai/marqo}, 2022.

\bibitem{ite_matsui_2018}
Y.~Matsui, Y.~Uchida, H.~J\'{e}gou, and S.~Satoh, ``A survey of product quantization,'' \emph{ITE Transactions on Media Technology and Applications}, vol.~6, no.~1, pp. 2--10, 2018.

\bibitem{Muja2009}
M.~Muja and D.~Lowe, ``Fast approximate nearest neighbors with automatic algorithm configuration.'' vol.~1, 01 2009, pp. 331--340.

\bibitem{pan2024survey}
J.~J. Pan, J.~Wang, and G.~Li, ``Survey of vector database management systems,'' \emph{The VLDB Journal}, vol.~33, no.~5, pp. 1591--1615, 2024.

\bibitem{pgvecto.rs}
pgvecto.rs, ``pgvecto.rs,'' \url{https://github.com/tensorchord/pgvecto.rs}, 2023.

\bibitem{Qdrant}
Qdrant, ``Qdrant,'' \url{https://github.com/qdrant/qdrant}, 2020.

\bibitem{Quan_2022}
\BIBentryALTinterwordspacing
L.~Quan, Q.~Guo, X.~Xie, S.~Chen, X.~Li, and Y.~Liu, ``Towards understanding the faults of javascript-based deep learning systems,'' in \emph{Proceedings of the 37th IEEE/ACM International Conference on Automated Software Engineering}, ser. ASE ’22.\hskip 1em plus 0.5em minus 0.4em\relax ACM, Oct. 2022, p. 1–13. [Online]. Available: \url{http://dx.doi.org/10.1145/3551349.3560427}
\BIBentrySTDinterwordspacing

\bibitem{Rigger_2020}
\BIBentryALTinterwordspacing
M.~Rigger and Z.~Su, ``Detecting optimization bugs in database engines via non-optimizing reference engine construction,'' in \emph{Proceedings of the 28th ACM Joint Meeting on European Software Engineering Conference and Symposium on the Foundations of Software Engineering}, ser. ESEC/FSE ’20.\hskip 1em plus 0.5em minus 0.4em\relax ACM, Nov. 2020, p. 1140–1152. [Online]. Available: \url{http://dx.doi.org/10.1145/3368089.3409710}
\BIBentrySTDinterwordspacing

\bibitem{silpa2008}
C.~Silpa-Anan and R.~Hartley, ``Optimised kd-trees for fast image descriptor matching,'' in \emph{2008 IEEE Conference on Computer Vision and Pattern Recognition}, 2008, pp. 1--8.

\bibitem{tagliabue2023vectorspacefinalfrontier}
\BIBentryALTinterwordspacing
J.~Tagliabue and C.~Greco, ``(vector) space is not the final frontier: Product search as program synthesis,'' 2023. [Online]. Available: \url{https://arxiv.org/abs/2304.11473}
\BIBentrySTDinterwordspacing

\bibitem{txtai}
txtai, ``txtai,'' \url{https://github.com/neuml/txtai}, 2020.

\bibitem{Vardanian_USearch_2023}
\BIBentryALTinterwordspacing
A.~Vardanian, ``{USearch by Unum Cloud},'' Oct. 2023. [Online]. Available: \url{https://github.com/unum-cloud/usearch}
\BIBentrySTDinterwordspacing

\bibitem{Vespa}
Vespa, ``Vespa,'' \url{https://github.com/vespa-engine/vespa}, 2016.

\bibitem{kappa}
S.~M. Vieira, U.~Kaymak, and J.~M.~C. Sousa, ``Cohen's kappa coefficient as a performance measure for feature selection,'' in \emph{International Conference on Fuzzy Systems}, 2010, pp. 1--8.

\bibitem{Voyager}
Voyager, ``Voyager,'' \url{https://github.com/spotify/voyager}, 2023.

\bibitem{2021milvus}
J.~Wang, X.~Yi, R.~Guo, H.~Jin, P.~Xu, S.~Li, X.~Wang, X.~Guo, C.~Li, X.~Xu \emph{et~al.}, ``Milvus: A purpose-built vector data management system,'' in \emph{Proceedings of the 2021 International Conference on Management of Data}, 2021, pp. 2614--2627.

\bibitem{Wang2021}
\BIBentryALTinterwordspacing
J.~Wang, X.~Yi, R.~Guo, H.~Jin, P.~Xu, S.~Li, X.~Wang, X.~Guo, C.~Li, X.~Xu, K.~Yu, Y.~Yuan, Y.~Zou, J.~Long, Y.~Cai, Z.~Li, Z.~Zhang, Y.~Mo, J.~Gu, R.~Jiang, Y.~Wei, and C.~Xie, ``Milvus: A purpose-built vector data management system,'' in \emph{Proceedings of the 2021 International Conference on Management of Data}, ser. SIGMOD '21.\hskip 1em plus 0.5em minus 0.4em\relax New York, NY, USA: Association for Computing Machinery, 2021, p. 2614–2627. [Online]. Available: \url{https://doi.org/10.1145/3448016.3457550}
\BIBentrySTDinterwordspacing

\bibitem{wang2017surveylearninghash}
\BIBentryALTinterwordspacing
J.~Wang, T.~Zhang, J.~Song, N.~Sebe, and H.~T. Shen, ``A survey on learning to hash,'' 2017. [Online]. Available: \url{https://arxiv.org/abs/1606.00185}
\BIBentrySTDinterwordspacing

\bibitem{wang2021-Survey}
\BIBentryALTinterwordspacing
M.~Wang, X.~Xu, Q.~Yue, and Y.~Wang, ``A comprehensive survey and experimental comparison of graph-based approximate nearest neighbor search,'' 2021. [Online]. Available: \url{https://arxiv.org/abs/2101.12631}
\BIBentrySTDinterwordspacing

\bibitem{wang2024llmsc}
\BIBentryALTinterwordspacing
S.~Wang, Y.~Zhao, X.~Hou, and H.~Wang, ``Large language model supply chain: {A} research agenda,'' \emph{CoRR}, vol. abs/2404.12736, 2024. [Online]. Available: \url{https://doi.org/10.48550/arXiv.2404.12736}
\BIBentrySTDinterwordspacing

\bibitem{wang2025sok}
\BIBentryALTinterwordspacing
S.~Wang, Y.~Zhao, Z.~Liu, Q.~Zou, and H.~Wang, ``Sok: Understanding vulnerabilities in the large language model supply chain,'' \emph{CoRR}, vol. abs/2502.12497, 2025. [Online]. Available: \url{https://doi.org/10.48550/arXiv.2502.12497}
\BIBentrySTDinterwordspacing

\bibitem{wang2025}
\BIBentryALTinterwordspacing
S.~Wang, Y.~Zhao, Y.~Xie, Z.~Liu, X.~Hou, Q.~Zou, and H.~Wang, ``Towards reliable vector database management systems: A software testing roadmap for 2030,'' 2025. [Online]. Available: \url{https://arxiv.org/abs/2502.20812}
\BIBentrySTDinterwordspacing

\bibitem{Weaviate}
Weaviate, ``Weaviate,'' \url{https://github.com/weaviate/weaviate}, 2016.

\bibitem{Wei2020}
\BIBentryALTinterwordspacing
C.~Wei, B.~Wu, S.~Wang, R.~Lou, C.~Zhan, F.~Li, and Y.~Cai, ``Analyticdb-v: a hybrid analytical engine towards query fusion for structured and unstructured data,'' \emph{Proc. VLDB Endow.}, vol.~13, no.~12, p. 3152–3165, Aug. 2020. [Online]. Available: \url{https://doi.org/10.14778/3415478.3415541}
\BIBentrySTDinterwordspacing

\bibitem{wu2022hqann}
\BIBentryALTinterwordspacing
W.~Wu, J.~He, Y.~Qiao, G.~Fu, L.~Liu, and J.~Yu, ``Hqann: Efficient and robust similarity search for hybrid queries with structured and unstructured constraints,'' 2022. [Online]. Available: \url{https://arxiv.org/abs/2207.07940}
\BIBentrySTDinterwordspacing

\bibitem{wu2025bhaktilightweightvectordatabase}
\BIBentryALTinterwordspacing
Z.~Wu, ``Bhakti: A lightweight vector database management system for endowing large language models with semantic search capabilities and memory,'' 2025. [Online]. Available: \url{https://arxiv.org/abs/2504.01553}
\BIBentrySTDinterwordspacing

\bibitem{yu2025mattersbridgingtopicssocial}
\BIBentryALTinterwordspacing
Q.~Yu, X.~Wang, S.~Liu, Y.~Bai, X.~Yang, X.~Wang, C.~Meng, S.~Wu, H.~Yang, H.~Xiao, X.~Li, F.~Yang, X.~Feng, L.~Hu, H.~Li, K.~Gai, and L.~Zou, ``Who you are matters: Bridging topics and social roles via llm-enhanced logical recommendation,'' 2025. [Online]. Available: \url{https://arxiv.org/abs/2505.10940}
\BIBentrySTDinterwordspacing

\end{thebibliography}

\end{document}